\documentclass[aip,cha,amsmath,reprint]{revtex4-1}
\usepackage{graphicx}
\usepackage{latexsym}
\usepackage{amsfonts}
\usepackage{color}

\begin{document}

\title{Nonreciprocal wave scattering on nonlinear string-coupled oscillators}
\date{\today}

\author{Stefano Lepri}
\email{stefano.lepri@isc.cnr.it}
\affiliation{Consiglio Nazionale delle Ricerche, Istituto dei Sistemi Complessi, 
via Madonna del Piano 10, I-50019 Sesto Fiorentino, Italy}
\affiliation{Istituto Nazionale di Fisica Nucleare, Sezione di Firenze, 
via G. Sansone 1, I-50019 Sesto Fiorentino, Italy}

\author{Arkady Pikovsky}
\affiliation{Department of Physics and Astronomy, University of Potsdam,
Karl-Liebknecht-Str 24/25, Potsdam, Germany}
\affiliation{Department of Control Theory, Nizhni Novgorod State 
University, Gagarin Av. 23,
606950, Nizhni Novgorod, Russia}

\begin{abstract}
 We study scattering of a periodic wave in a string on  two lumped oscillators attached to it. The equations can be represented as a driven (by the incident wave) dissipative (due to radiation losses) system of delay differential equations of neutral type. Nonlinearity of oscillators makes
the scattering non-reciprocal: the same wave is transmitted differently in two
directions. Periodic regimes of scattering are analyzed approximately, using amplitude
equation approach. We show that this setup can act as a \textit{nonreciprocal modulator} via
Hopf bifurcations of the steady solutions. Numerical simulations of the full system reveal nontrivial regimes of quasiperiodic and chaotic scattering. Moreover, a regime of a \textit{``chaotic diode''}, where transmission is periodic in one direction and chaotic in the opposite one, is reported.
\end{abstract}

\pacs{05.60.-k 05.70.Ln 44.10.+i}

\maketitle

\begin{quotation}
One of the mostly general results of the linear wave theory 
is the reciprocity theorem, established in works of Rayleigh, Helmholtz and Lorentz.
For the one-dimensional wave scattering it means the symmetry of the scattering matrix, so that transmission
in both direction is the same. While in linear systems violations of reciprocity require
violations of time-reversal symmetry, in nonlinear wave propagation reciprocity does not hold.
In particular, scattering of linear waves on nonlinear objects may operate as a ``wave diode'', with different 
transmission properties in both directions. Here we consider a simple model of scattering of linear waves on two 
lumped nonlinear oscillators. If one neglects dispersion and dissipation in the medium and in the oscillators, the 
equations can be reduced to a system of delay-differential equations. We demonstrate in this paper different regimes 
of reciprocity violations. In the simplest case transmissions in both directions are different, while the waves 
remain periodic. We observe also more complex regimes, where reflected and transmitted waves are chaotic and 
different. Probably, mostly nontrivial regime reported is that of ``chaotic diode'': a periodic wave sent to the 
scatterer in one direction remains periodic, while when the same wave is sent in another direction, transmitted and 
reflected waves are chaotic. 
\end{quotation}

\section{Introduction}

Understanding the way in which nonlinearity affects wave propagation is one of
the  basic issues in many different domains such as nonlinear optics,
acoustics, electronics and fluid dynamics. 
A related challenging goal is the control of wave
energy flow using fully nonlinear features.

The most elementary form of control would be to devise a ``wave diode'' in which
some input energy is transmitted differently along two opposite propagation
directions. As it is known, this is forbidden in a linear, time-reversal 
symmetric system, by virtue of the reciprocity theorem
\cite{Rayleigh}. The standard way to circumvent this limit is to break the
time-reversal  symmetry by applying a magnetic field, as done, for instance, in
the case of optical isolators.  An entirely alternative possibility is instead
to consider \textit{nonlinear} effects. At least in principle, this option would
offer novel possibilities of propagation control based on intrinsic material
properties  rather than on an external field. A general critical discussion  of
those issues can be found in Ref.~\onlinecite{Maznev2012}. 

The idea of exploiting nonlinear effects has been pursued in different 
contexts. In the domain of lattice dynamics,  asymmetric phonon transmission
through a nonlinear layer between two very dissimilar crystals has been
demonstrated in Ref.~\onlinecite{Kosevich1995}.  Other concrete examples are offered
by nonlinear phononic media \cite{Liang09,Liang2010} and the propagation of
acoustic pulses through granular systems \cite{Nesterenko05,Boechler2011}.
Nonlinear optics is also a versatile playground 
as exemplified by the so-called all-optical diode 
\cite{Scalora94,Tocci95,Gallo01}. In particular, in Ref.~\onlinecite{Maes:06} symmetry-breaking in two nonlinear microcavities has been described.  Other proposals include left-handed metamaterials
\cite{Feise05}, quasiperiodic systems \cite{Biancalana08},  coupled nonlinear
cavities \cite{Grigoriev2011} or $\mathcal{PT-}$symmetric waveguides
\cite{Ramezani2010,D'Ambroise2012,Bender2013} and transmission lines \cite{Tao2012}. 
Extensions to the quantum systems 
\cite{Roy2010,Mascarenhas2014} and nonlinear 
Aharanov-Bohm rings \cite{Li2014} have been also considered.

Despite the variety of physical contexts, the basic underlying  rectification
mechanisms rely on nonlinear phenomena as, for instance,  second-harmonic
generation in photonic \cite{Konotop02} or phononic crystals \cite{Liang09}, or
bifurcations \cite{Boechler2011}. In those examples the rectification depends
on  whether some harmonic (or subharmonic) of the fundamental wave is
transmitted or not. As discussed in Ref.~\onlinecite{Maznev2012} a more strict  
operating condition would be that the transmitted  power at the
\textit{same frequency and incident amplitude} would be sensibly different in
the two opposite propagation directions. Nonlinear resonances have been 
proved to be effective in achieving this \cite{Lepri2011} (see also
Ref.~\onlinecite{Xu2014} where Fano resonances have been considered).

The above issues are conveniently studied as a scattering problem  i.e. by
seeking for wave solutions impinging on a nonlinear impurity. In a
one-dimensional geometry such solutions can be found by simple methods  like the
transfer-matrix approach (see \cite{Lepri2011} and  references therein). Once
the solutions are known one natural question is the assessment of their
dynamical stability and bifurcations.  This question has been investigated only
to a limited extent \cite{Malomed93,Miroshnichenko2009}. More recently, it has
been shown that   scattering states in the presence of (generally complex)
impurities  typically display oscillatory instabilities \cite{DAmbroise2013}
that may  results in the creation of stable quasiperiodic, nonreciprocal
solutions \cite{Lepri2014}. Those can be seen as a superposition of an 
extended wave with a nonlinear defect mode oscillating at a different frequency.  
It can be envisaged that more complex dynamical regimes may be observed and that
this will affect the overall performance of any device that one may wish to
realize in practice.

In the present paper we introduce a simple model for a scalar wave field
interacting with two different local nonlinear elements. It is a generalization
of the system introduced in Ref.~\onlinecite{Pikovsky1993} as a simple example of
chaotic wave scattering, where only one local nonlinear oscillator coupled to a
wave medium was considered. Clearly, with one lumped oscillator the 
scattering is fully reciprocal, although non-trivial.
The model we consider belongs to a class of wave systems with local nonlinearity. In the case of dispersive waves (e.g. in a lattice~\cite{Brazhnyi-Malomed-11,Brazhnyi2014277} or with a periodic
background potential~\cite{PhysRevA.83.033828}) such a system
can possess localized solutions (breathers); a similar
situation occurs in a Schr\"odinger equation with local nonlinearity that
creates local pseudopotential well where wave is localized~\cite{PhysRevA.78.053601}. 
We consider here non-dispersive waves, this system does not possess localized solutions.

As it will be shown in Section \ref{sec:model},  our model
can be reformulated as a delay-differential equations and thus admits a very
rich dynamics depending of the  relation between its relevant time scales.
Indeed, complex input-output responses can be easily achieved, including
quasiperiodic and chaotic ones. In Section \ref{sec:amp} we start the analysis 
of the system by considering the case of weak coupling between the string and
the oscillators. This limiting case can be treated  by means of approximate
amplitude equations.  In Section \ref{sec:gen} we turn to the more general case
in  which there is no sharp separation among timescales and the  system can only
be
treated by direct numerical integration of the  full set of equation. Here the
dynamics is considerably more  complex, leading to high-dimensional and possibly
chaotic motion.

\section{Basic equations}
\label{sec:model}

\begin{figure}[ht]
\begin{center}
\includegraphics[width=0.45\textwidth,clip]{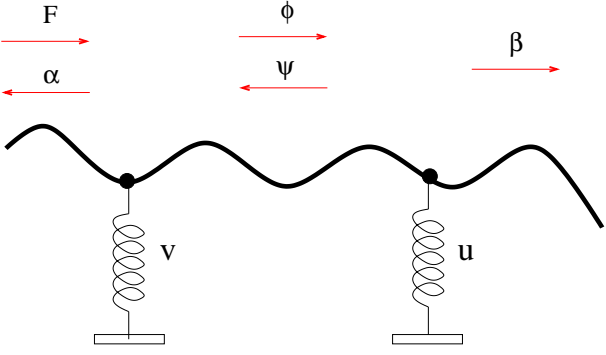}
\caption{The model.}
\label{fig:model}
\end{center}
\end{figure}

The model is inspired by Ref.~\onlinecite{Pikovsky1993} and is schematically depicted in Fig.~\ref{fig:model}. 
It amounts to two lumped, undamped oscillators $v(t)$ and $u(t)$ attached 
to an elastic string at points $-L/2$ and $L/2$ 
correspondingly. The equations of motion for the oscillators are
\begin{gather}
m_1\ddot v+V(v)=S\left(\frac{\partial y^0}{\partial x}-\frac{\partial y^-}{\partial x}\right)_{x=-L/2}\\
m_2\ddot u+U(u)=S\left(\frac{\partial y^+}{\partial x}-\frac{\partial y^0}{\partial x}\right)_{x=L/2}
\end{gather}
Here $y^-(t,x)$, $y^0(t,x)$, and  $y^+(t,x)$ denote the string 
displacement in the domains
$[-\infty,-L/2)$, $(-L/2,L/2)$ and $(L/2,\infty]$ respectively;
$U$ and $V$ are the local forces acting on the two oscillator that we 
assume to be different to break the mirror symmetry of the system
around $x=0$. The string obeys the equation of motion
\[
\frac{\partial^2 y}{\partial t^2}-c^2\frac{\partial^2 y}{\partial x^2}=0\;,\qquad
c^2=\frac{S}{\rho}\;,
\]
where $S$ is the tension and $\rho$ is the mass density.
The energy density of the wave is
\[
\mathcal{E}=\frac{1}{2}S\left(\frac{\partial y}{\partial x}\right)^2
+ \frac{1}{2}\rho \left(\frac{\partial y}{\partial t}\right)^2
\]
and the energy conservation reads
\begin{gather}
\frac{\partial }{\partial t}\mathcal{E}+\frac{\partial }{\partial x}J=0\\
J=-S\frac{\partial y}{\partial x}\frac{\partial y}{\partial t}
\end{gather}
with $J$ being the energy flux.
We represent the string field as
\begin{eqnarray*}
&&\text{Incident wave: } F(t-(x+L/2)/c) \text{ for } x<-L/2\\
&&\text{reflected wave: } \alpha(x/c+t) \text{ for } x<-L/2\\
&&\text{transmitted wave: } \beta(t-x/c) \text{ for } x>L/2\\
&&\text{``interaction'' waves } \phi(t-x/c),\;\psi(t+x/c) \\
&& \text{ for } -L/2<x<L/2
\end{eqnarray*}

In the Appendix we show that the problem can be reduced to a coupled system
of delay-differential equations for the variables describing the two oscillators. 
The possibility of such a reduction heavily relies on the non-dispersive, non-dissipative nature of wave propagation along the string. 
In the case of dispersion and dissipation, one would
obtain integro-differential equations that are very hard to investigate.

For convenience, 
we introduce the time scale according to some frequency $\Omega$, so that the new dimensionless time will be $\tau=\Omega t$.
In terms of the dimensionless time delay $T=\Omega L/2c$ and dimensionless coupling parameter $a=\frac{S}{mc\Omega}$,
the system of equations reads
\begin{gather}
\frac{d^2 v}{d\tau^2}+2a\frac{dv}{d\tau}+\frac{V(v)}{m\Omega^2}=2a\dot F(\Omega^{-1}\tau)+2a\dot \psi(\tau-T)\label{eq:dde1}\\
\frac{d^2 u}{d\tau^2}+2a\frac{du}{d\tau}+\frac{U(u)}{m\Omega^2}=2a\dot\phi(\tau-T))\label{eq:dde2}\\
\dot\psi(\tau)=\dot u(\tau-T)-\dot\phi(\tau-2T)\label{eq:dde3}\\
\dot\phi(\tau)=\dot v(\tau-T)-\dot\psi(\tau-2T)
\label{eq:dde4}
\end{gather}
From the system solution we can compute reflected and transmitted waves as
\begin{equation}
\alpha(t-L/2c)=v(t)-F(t);\qquad \beta(t-L/2c)=u(t).
\end{equation}
Moreover, one can evaluate the reflected and transmitted fluxes as
\begin{gather}
J_{refl}=-\sqrt{S\rho} \, \dot\alpha^2;\qquad J_{trans}=\sqrt{S\rho}\, \dot\beta^2.
\end{gather}

It should be remarked that the system (\ref{eq:dde1}-\ref{eq:dde4}) differs from
standard delayed  dynamical systems (like Ikeda, Mackey-Glass etc.) in several
respects.  Indeed, one typically has only terms delayed by $T$ while here we
have also a reflected components delayed by $2T$. Moreover, and more
importantly, the delayed coupling occurs via the derivatives of the
variables. This is referred to as  ``neutral type" of delay-differential
equation \cite{Hale1993,Bainov1991}.
Such equations also naturally appear in electrical networks, 
where lumped elements are connected with lossless transmission lines \cite{Miranker1961,Brayton1968}  
that, in fact, is the setup equivalent to the mechanical one of Fig.~\ref{fig:model}. 

Noteworthy, the system (\ref{eq:dde1}-\ref{eq:dde4}) is \textit{dissipative}.
This is the radiation losses, as the only sink of energy is due to reflected and
transmitted waves. The dissipation parameter is the coupling parameter $a$.

\section{Amplitude equations and their analysis}
\label{sec:amp}

Let us consider Eqs.(\ref{eq:dde1}-\ref{eq:dde4}) and set units such that $\Omega=m=c=1$.
Furthermore, we specialize to the case of a periodic wave 
forcing $F(t) = F e^{i\omega t} + c.c. $. The dynamics is thus 
characterized by three main timescales, $T$, $1/a$, $1/\omega$.
In this Section we first focus on the case of weak coupling whereby
$1/a$ is much larger than both $T$ and $1/\omega$.
For definiteness, we consider forces of the form
\begin{equation}
U(u) = \omega_1^2 u + k_1 u^3\;, \quad V(v) = \omega_2^2 v + k_2 v^3\;,
\label{eq:forces}
\end{equation}
and distinguish three distinct regimes where the system equations 
can be simplified by suitable approximations. 

\subsection{1:1 resonance}

Let us first consider the case in which $\omega \sim \omega_1 \sim \omega_2$.
We look for an expansion in slowly varying amplitudes (assuming weak dissipation $a$):
\begin{eqnarray*}
&v(t) = A e^{i\omega t} + c.c. &\quad \dot v(t) = i\omega A e^{i\omega t} + c.c. \\
&u(t) = B e^{i\omega t} + c.c. &\quad \dot u(t) = i\omega B e^{i\omega t} + c.c. \\
&\phi(t) = \Phi e^{i\omega t} + c.c. 
&\quad \dot \phi(t) = i\omega \Phi e^{i\omega t}+ c.c. \\
&\psi(t) = \Psi e^{i\omega t} + c.c. &\quad \dot \psi(t) = i\omega \Psi e^{i\omega t} + c.c.
\end{eqnarray*}
In the same approximation the transmitted intensity is proportional 
to $|B|^2$. Making use of the rotating wave approximation, we 
neglect higher-order harmonics i.e. 
\[
v^3 \approx 3|A|^2A e^{i\omega t} + c.c.;
\quad u^3 \approx 3|B|^2B e^{i\omega t} + c.c.
\]
Equating terms proportional to $\sim e^{i\omega t}$ and keeping 
the lowest order in $a$ in the second-order derivatives we obtain
\begin{eqnarray}
&&i\dot A + (\Delta_1 + ia+ \gamma_1 |A|^2)A = 
ia \left( F + \Psi(t-T) e^{-i\omega T}\right)   \nonumber \\
&&i\dot B + (\Delta_2 +ia + \gamma_2 |B|^2)B = ia \Phi(t-T) e^{-i\omega T} \nonumber\\
&&\Psi(t) = B(t-T) e^{-i\omega T} - \Phi(t-2T) e^{-2i\omega T}\\  
&&\Phi(t) = A(t-T) e^{-i\omega T} - \Psi(t-2T) e^{-2i\omega T} \nonumber
\label{eq:ampeq}
\end{eqnarray}
where we have defined the detunings 
\[
\Delta_1 = \frac{\omega_1^2-\omega^2}{2\omega} \approx \omega_1-\omega;\quad 
\Delta_2 = \frac{\omega_2^2-\omega^2}{2\omega} \approx \omega_2-\omega
\]
and the new nonlinearity parameters $\gamma_{1,2}=3k_{1,2}/2\omega$.

The steady state solutions are thus given by the system of transcendental
equations 
\begin{equation}
\begin{aligned}
&& (\Delta_1 +ia + \gamma_1 |A|^2)A = ia \left( F + \Psi e^{-i\omega T}\right)   \\
&& (\Delta_2 +ia + \gamma_2 |B|^2)B = ia \Phi e^{-i\omega T}\\
&&\Psi = B e^{-i\omega T} - \Phi e^{-2i\omega T}\\  
&&\Phi = A e^{-i\omega T} - \Psi e^{-2i\omega T}
\end{aligned}
\label{eq:stst}
\end{equation}
Solving the last two equations and substituting in the first two yields
\begin{eqnarray*}
&&\Psi = 
\frac{-Ae^{-i\omega T}+B e^{i\omega T}}{e^{2i\omega T}-e^{-2i\omega T}}\\  
&&\Phi = 
\frac{Ae^{i\omega T}-Be^{-i\omega T}}{e^{2i\omega T}-e^{-2i\omega T}}\\
&& (\Delta_1 +ia + \gamma_1 |A|^2)A = \\ 
&& = ia \left( F + 
\frac{-Ae^{-i\omega T}+B e^{i\omega T}}{e^{2i\omega T}-e^{-2i\omega T}}
e^{-i\omega T}\right)   \\
&& (\Delta_2 +ia + \gamma_2 |B|^2)B = ia
\frac{Ae^{i\omega T}-B e^{-i\omega T}}{e^{2i\omega T}-e^{-2i\omega T}}
e^{-i\omega T}
\end{eqnarray*}
Note that this solution runs into troubles when $e^{-2i\omega T}=1$,
since then the system is 
undetermined (a ``small denominator'' problem). Using terminology from optics, 
this corresponds to the Fabry-Perot resonances 
$\omega_n = \frac{\pi n}{2T}$, $n$ integer, of the modes of the ``cavity'' represented
by the portion of the string comprised between the oscillators.

Away from such resonances equations are solved by introducing the 
amplitudes and phase shifts as $A=|A|e^{i\theta}$, 
$B=|B|e^{i(\theta+\rho)}$.
Eliminating $\theta$ and $\rho$ we obtain
\begin{eqnarray}
&& \left[C|A|^2-D|B|^2\right]^2 +\frac{a^2}{4}\left[ |A|^2+|B|^2\right]^2 
=a^2|A|^2F^2 \nonumber\\
&& \left[D^2+\frac{a^2}{4}\right]|B|^2 = \frac{a^2}{4}
\frac{|A|^2}{\sin^2 2\omega T}
\label{mstable}
\end{eqnarray}
where
\begin{eqnarray*}
&& C\equiv \Delta_1 +\gamma_1 |A|^2+\frac{a}{2} \cot 2\omega T\\
&& D\equiv \Delta_2 +\gamma_2 |B|^2+\frac{a}{2} \cot 2\omega T
\end{eqnarray*}

\begin{figure}[h]
\begin{center}
\includegraphics[width=0.5\textwidth,clip]{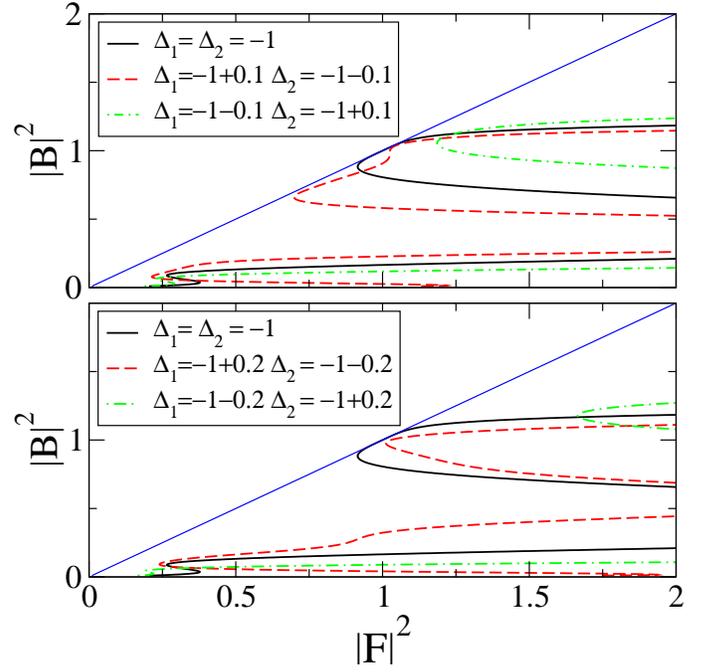}
\caption{Input-output curves computed from Eq.~(\ref{mstable})
for $a^2=0.4$, $\gamma_1=\gamma_2=1$, 
$\omega T=\pi/4$
and different detuning values. In both panels the solid black line corresponds
to the symmetric case $\Delta_{1,2}=1$ while dashed and dot-dashed lines
are respectively for 
and $\Delta_{1,2}=-1(1\pm 0.1)$ (upper panel) and 
$\Delta_{1,2}=-1(1\pm 0.2)$ (lower panel)
i.e. the left-to-right and right-to-left transmission for unequal oscillators. 
The thin solid line is the bisectrix. 
Note that in the second case for $F\approx 1$ there is basically no transmission
in one of the two directions.
}
\label{f:inout}
\end{center}
\end{figure}

In the symmetric case (i.e. when the oscillators are equal) $\Delta_1=\Delta_2=\Delta$, 
$\gamma_1=\gamma_2=\gamma$, 
perfectly transmitted solutions $|A|=|B|=F$ exist for 
$C=D=\pm \frac{a}{2} \cot 2\omega T$ i.e. for
\[
\Delta +\gamma F^2=0;\quad  \Delta +\gamma F^2= a\, \cot 2\omega T\quad.
\]
The last equations determine the nonlinear resonances of the system:
whenever such solutions exist, a multistable regime is expected where 
asymmetric propagation should set in \cite{Lepri2011} .
This is confirmed in Fig.\ref{f:inout} where we plot $|B|^2$ versus $F^2$ 
in the bistable regime 
and compare the symmetric case with two ones in which $\Delta_1\neq\Delta_2$.
The forward (resp. backward) case corresponds to an input applied to 
the first (resp. second) oscillator. This is obviously equivalent to
compare solutions of (\ref{eq:stst}) whereby the two oscillator are exchanged.
As it is seen, there are regions close to the nonlinear resonance 
in which the same input can be transmitted very differently \cite{Lepri2011}.
In the case of the lower panel of Fig.\ref{f:inout}, transmission in one direction 
is actually almost suppressed.

To conclude this Subsection, we comment on the dynamics close to the 
Fabry-Perot resonances. To this aim  
we let $\omega=\omega_n+\epsilon$ with $\epsilon$ being a smallness 
parameter such that $e^{\pm 2i\omega T}\approx 1 \pm 2i\epsilon $
and assume a perturbative expansion
\[
A=A_1\epsilon + A_2 \epsilon^2\ldots; \quad 
B=B_1\epsilon + B_2 \epsilon^2\ldots;\quad 
F=F_1\epsilon
\]
(the last is just a rescaling of the force).
Substituting and equating the leading order terms 
\begin{eqnarray*}
O(1):&&\quad A_1 = B_1 \\
O(\epsilon):&& (\Delta_1 +ia)A_1=iaF_1+\frac{a}{4}(-A_2+B_2);\\
&& (\Delta_2 +ia)A_1=\frac{a}{4}(A_2-B_2)
\end{eqnarray*}
From which we find the solution up to corrections $O(\epsilon)$: 
\[
A=B= \frac{iaF}{\Delta_1 +\Delta_2 +2ia}; \quad
\frac{|B|^2}{|F|^2}=\frac{a^2}{(\Delta_1 +\Delta_2)^2 +4a^2}
\]
Note that in this limit the nonlinear terms are irrelevant 
and transmission coefficient is therefore symmetric with
respect to the exchange of the two oscillators. So we do
not expect sizable reciprocity violations close to resonances.

\subsection{Higher-order resonances}

As mentioned in the Introduction, we are mostly interested in the case of a
transmitted wave having mostly the same frequency as the input one.  For completeness,
we briefly touch on the problem of higher-order resonances which can be studied
with a similar approach. Let us consider for instance the case of a 1:3
resonance, namely  the one in which $\omega \sim \omega_1 \sim \omega_2/3$.
Conceptually, this corresponds to experimentally relevant situations in which
the rectification is induced by excitation of higher-order harmonics 
\cite{Liang09,Tao2012}.

The mechanism at work here is the following: the incident wave weakly excites 
the third harmonic of the first oscillator. The latter is in resonance with 
the second and can be transmitted. On the other hand, excitation of the second 
oscillator is negligible since almost no power can be transferred. 
This suggests looking for solutions of the form
\begin{eqnarray*}
&v(t) = A e^{i\omega t} + a A_3 e^{3i\omega t}+ c.c. \quad 
& \dot v(t) = i\omega A e^{i\omega t} + c.c. \\
&u(t) = a  B e^{3i\omega t} + c.c. & \dot u(t) = 3ia\omega B e^{3i\omega t} + c.c. 
\end{eqnarray*}
with $\phi$ and $\psi$ having the same form as in previous Subsection.
The coupling between oscillators should thus occur through the third-harmonic amplitude 
$A_3$. This means that the asymmetry of transmission should be pretty weak,  
of order $a^2$, and thus not very effective. 

\subsection{Small delay limit}

Consider the case in which $a\ll\omega \sim 1/T$ but 
$\omega T \to const$. In this limit we can neglect 
the delay in the Eqs.~\eqref{eq:ampeq}
(up to terms of order $O(a^2)$).
It means that the retardation effects enter only through
phase shifts. 
Expressing $\Psi$, $\Phi$ as a function of $A,B$ in the last two Eqs.~(\ref{eq:ampeq}),
we get
\begin{eqnarray}
&&i\dot A + (\delta_1 + \frac{ia}{2}+ \gamma_1 |A|^2)A = 
ia  F + \kappa B  \nonumber \\
&&i\dot B + (\delta_2 +\frac{ia}{2} + \gamma_2 |B|^2)B = \kappa A
\label{nodelay}
\end{eqnarray}
where we have introduced the new detunings and coupling
\begin{equation}
 \delta_{1,2}\equiv \Delta_{1,2} + \frac{a}{2} \cot 2\omega T; \quad 
 \kappa \equiv \frac{a}{2\sin 2\omega T}.
\end{equation}
Before proceeding further we note that these equations resemble the
ones obtained in Ref. \onlinecite{PhysRevB.81.115128} for a photonic Fabry-Perot 
resonator coupled with two off-channel defects.

We performed some numerical experiments with these simplified equations (in rescaled units
in which $a=1$ was set).The generic findings are:
\begin{enumerate}
\item For a given external input $F$, the dynamics approaches 
a fixed point or a limit cycle, neither quasiperiodicity nor chaos is observed.
\item The nonreciprocal behavior manifests itself in all the possible combinations
of constant output in both directions or constant in one direction 
and periodic in the other. As
this would correspond to a modulation of output in the original model,
we may term this as a \textit{nonreciprocal modulator}.
\item The underlying Hopf bifurcations are typically subcritical
when the nonlinearities have the same sign and supercritical otherwise.
\end{enumerate}

\begin{figure*}
\includegraphics[width=0.7\textwidth,clip]{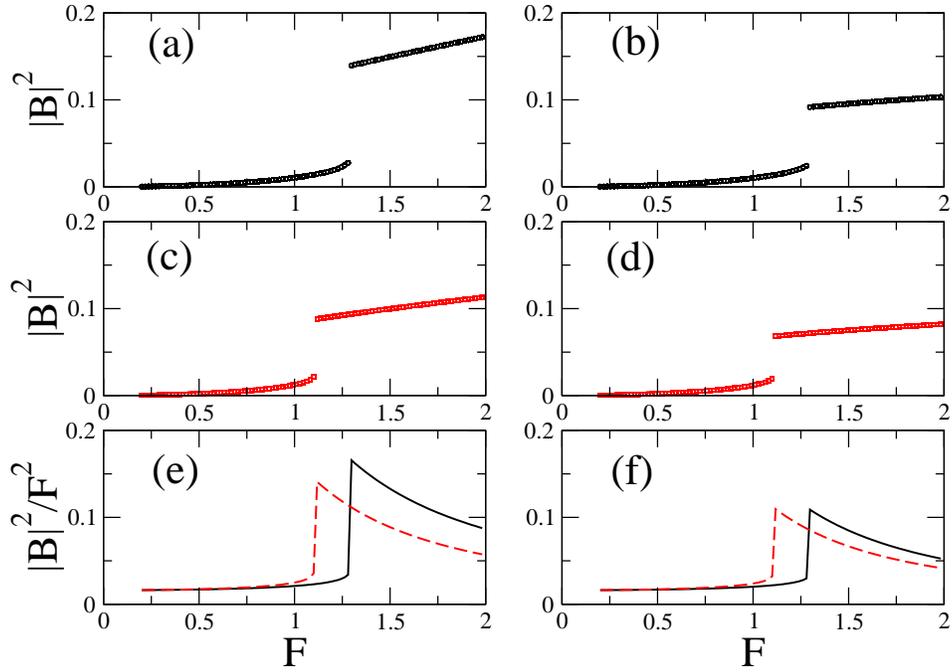}
\caption{Simulations of Eqs.~(\ref{nodelay}): input-output curves (a-d) 
and transmission coefficients (e,f) 
for $\Delta_{1,2}=-2.5(1\pm0.05)$ and $\omega T=0.5$. Left panels:
$\gamma_1=\gamma_2=1$
right panels $\gamma_1=-\gamma_2=1$. 
Panels (a,b) and solid black lines in panels (d,f) refers to the 
forward propagation; (c,d) and dashed red lines in (d,f) to the 
backward one; input amplitude $F$ is 
increased in steps from the  lowest value. }
\label{f:sim}
\end{figure*}

\begin{figure*}
\includegraphics[width=0.7\textwidth,clip]{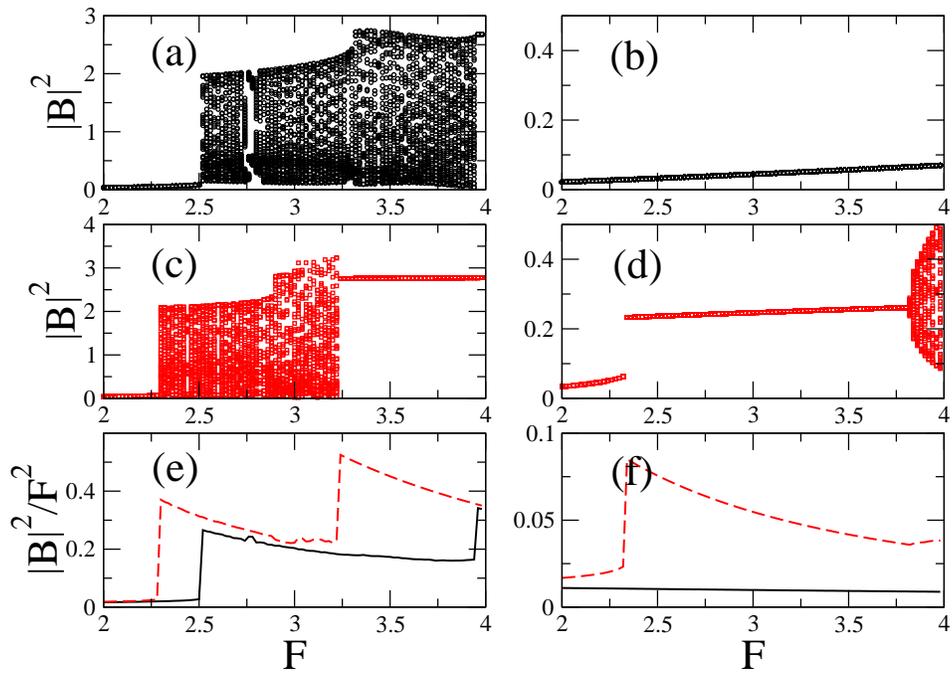}
\caption{Simulations of Eqs.~(\ref{nodelay}): 
same as in previous figure with $\omega T=1.4$.
For the oscillating amplitude regimes, in panels (e,f)
the time-averaged amplitudes are reported. }
\label{f:sim2}
\end{figure*}

The results are exemplified in Figs.~\ref{f:sim} and \ref{f:sim2}.
For instance, panels Figs.~\ref{f:sim2}(a) and (c) display a case of 
a nonreciprocal modulation. Indeed, the 
output in the forward direction is modulated periodically for amplitudes
larger than $F=2.5$ where a subcritical Hopf bifurcation sets in.
On the contrary, the output in the backward direction remains periodic
in the same ranges of input amplitudes.

\clearpage
\section{General case}
\label{sec:gen}

Here we discuss the case where no separation of time scales occurs and
we have to integrate the full system Fig.~\ref{fig:model} numerically. We report
on the results of numerical simulations of (\ref{eq:dde1}-\ref{eq:dde4}) 
for $T=8$. The potentials of the two point oscillators
are taken in form \eqref{eq:forces}. The system has many parameters, main of them are the eigenfrequencies of the oscillators. In most
of the numerical results below we use $\omega_1=1.2$, $\omega_2=1$, and $k_1=k_2=1$.

\begin{figure}
\includegraphics[width=\columnwidth]{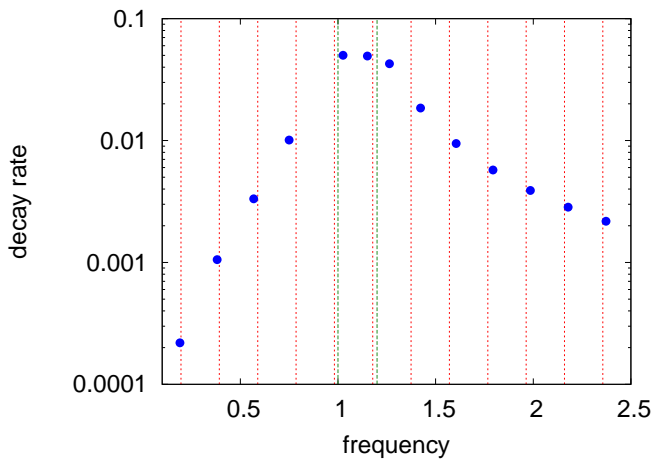}
\caption{Eigenmodes (blue filled circles) for $\omega_1=1$, $\omega_2=1.2$ and $a=0.25$. 
Red dotted lines show
Fabry-Perot resonances $\omega_k=k\pi/(2T)$, green dashed lines 
show the oscillator frequencies $\omega_{1,2}$.}
\label{fig:linmodes}
\end{figure}

To get some insight on the dynamics we first analyze the system
(\ref{eq:dde1}-\ref{eq:dde4},\ref{eq:forces}) linearized around the trivial fixed point. We report on
resulting eigenvalue spectrum in Fig.~\ref{fig:linmodes}.  One can see that
while eigenmodes (the Fabry-Perot modes)
with frequencies close to that of the oscillators have large
decay rates, those with small and large frequencies have very low decay rates.
This is a well-known property of hyperbolic systems, and correspondingly of
delay systems of neutral type like (\ref{eq:dde1}-\ref{eq:dde4}). Large (in
fact, infinite) number of nearly neutral modes makes many methods of numerical
analysis hardly applicable. To avoid excitation of such
high-frequency modes, we nearly
adiabatically switched on the external field in the study of scattering of the wave on the
oscillators. 

The main parameters that we change in the study of wave propagation, 
are the frequency $\omega$ and the amplitude
$A$ of the incoming wave, as we choose in \eqref{eq:dde1} $\dot F=\omega
A\cos(\omega \tau)$. For each amplitude, we focus on violations of reciprocity.
Given initially an empty system, we send a wave with the amplitude slowly
growing from zero to the maximal value, after which this amplitude remains
constant. After transients, we calculate the average transmitted and reflected
power; furthermore, correlation properties of the transmitted and reflected
waves have been analyzed.  

\begin{figure}
\includegraphics[angle=270,width=\columnwidth]{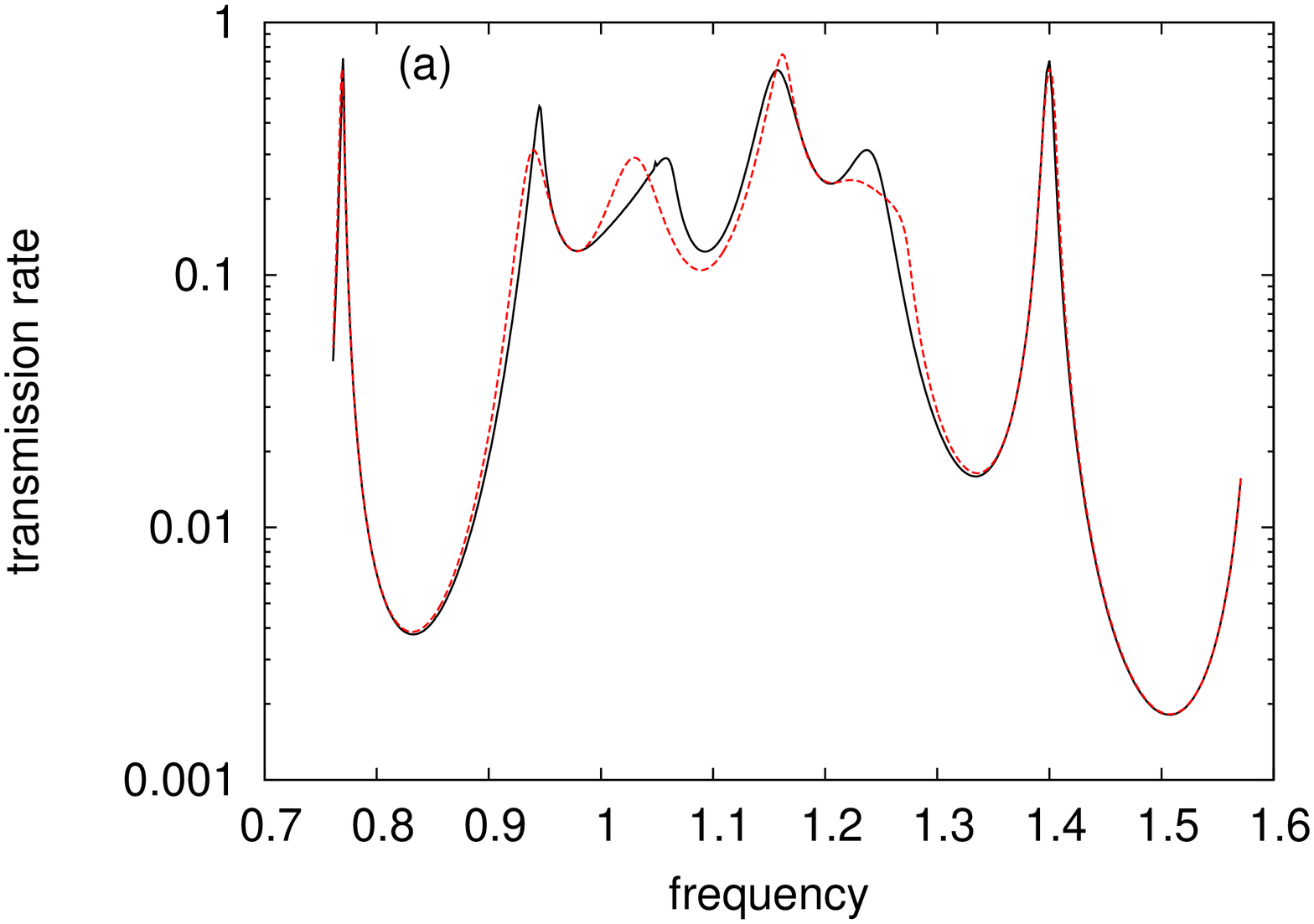}
\includegraphics[angle=270,width=\columnwidth]{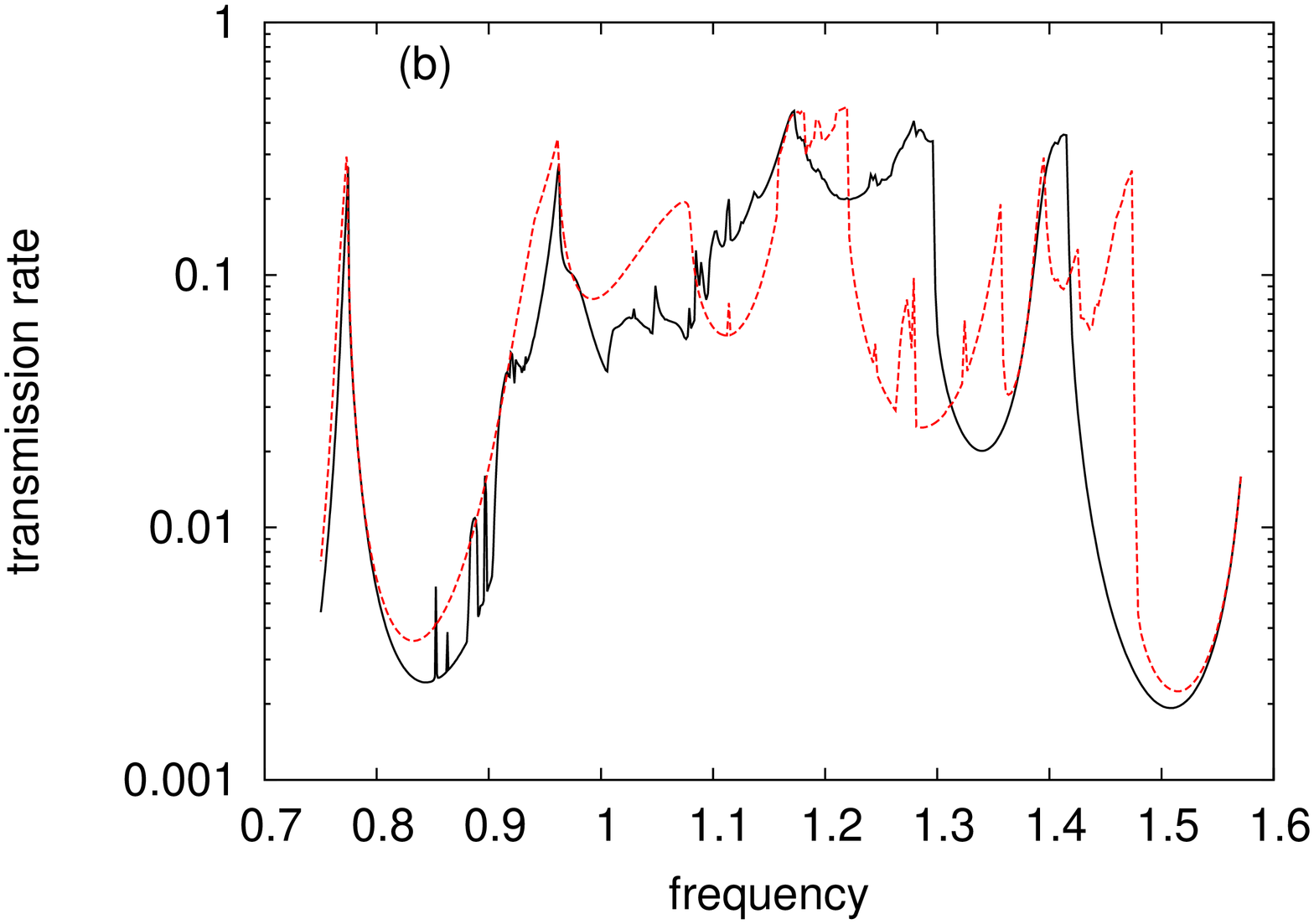}
\caption{Transmission coefficients in dependence on the frequency of the incoming wave,
for $a=0.1$ and two amplitudes: (a) $A=0.2$ and (b) $A=1$. Solid black line: left-to-right, dashed red line: right-to-left.}
\label{fig:amp0210}
\end{figure}

\begin{figure}
\includegraphics[angle=270,width=\columnwidth]{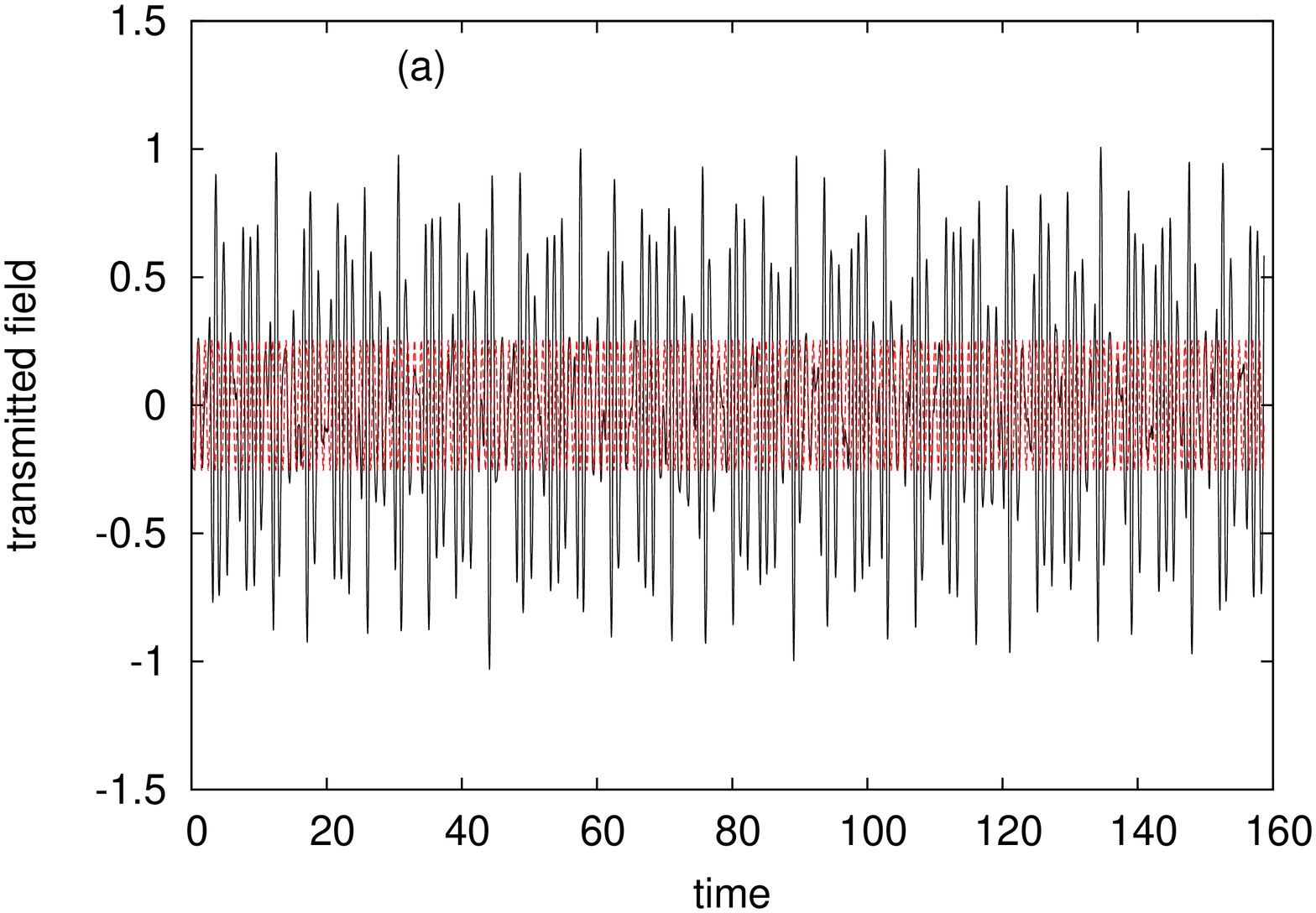}
\includegraphics[angle=270,width=\columnwidth]{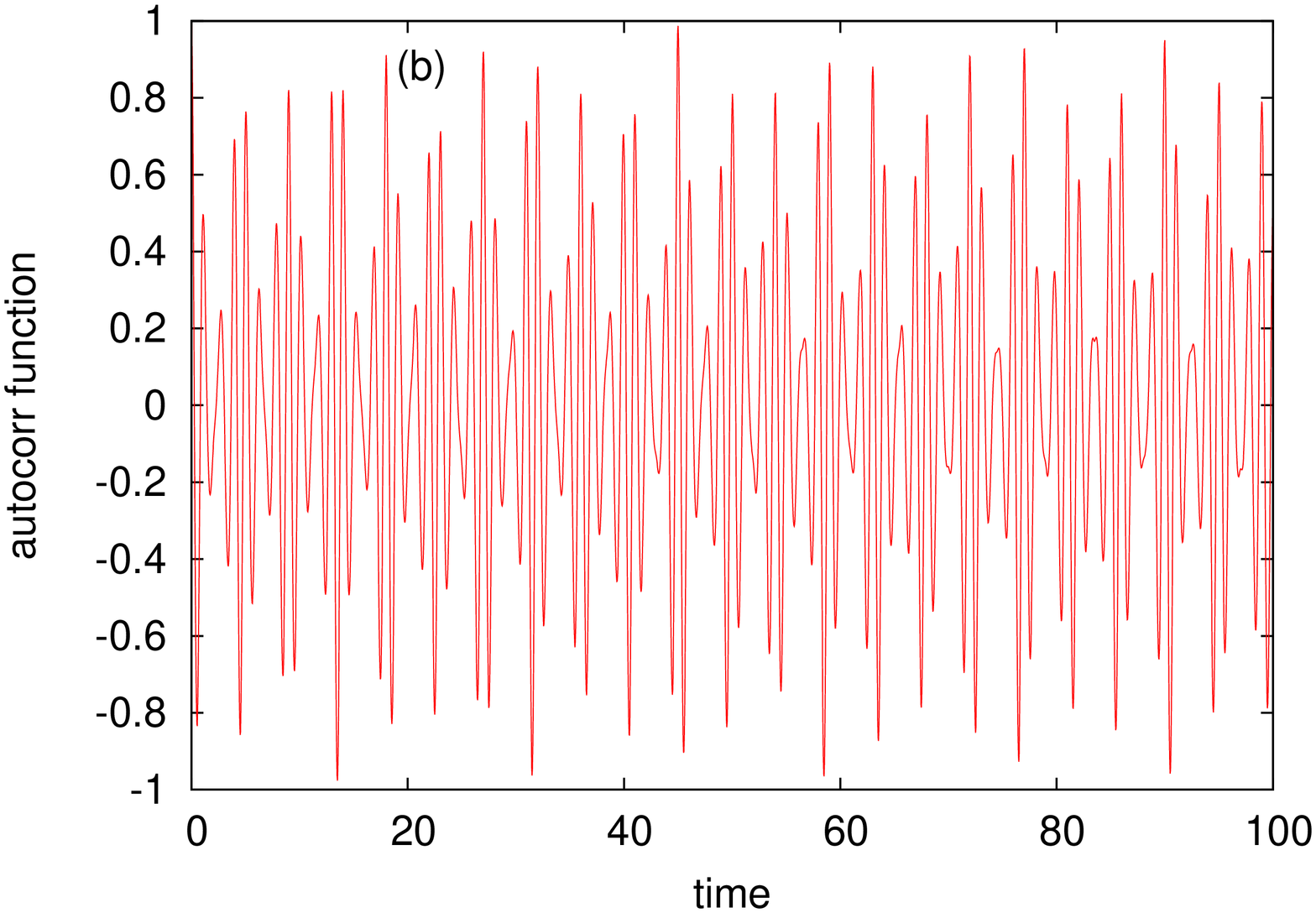}
\caption{(a) Transmitted waves from left to right (black) and from right to left (red).
(b) Autocorrelation function of the black field in panel (a). Time axis is
in units of the period of the incident wave.}
\label{fig:field}
\end{figure}

As one can expect, for small amplitudes of the incoming wave the system is fully
reciprocal, and  we illustrate first deviations from this in
Fig.~\ref{fig:amp0210}(a) where the results for a relatively small amplitude
$A=0.2$ are shown. Here nonlinear effects are maximal in the range of
frequencies close to that of oscillators, while outside of the range
$0.9\lesssim \omega\lesssim 1.3$ the transmission rates in both directions 
follow the
structure of linear modes. Non-reciprocity is much stronger expressed at a larger
amplitude $A=1$ (panel (b)).  Moreover, here the complexity of the field is
rather different for the two ways of propagation. We illustrate this in
fig.~\ref{fig:field}, where we show transmitted waves for $A=1$ and
$\omega=1.2467$. While the wave transmitted in one direction is periodic, 
the wave transmitted in the other direction has a more complex form. Detailed
analysis of the autocorrelation function shows however, that the correlations 
do not
decay but the whole process appears quasiperiodic (at the level 
of our numerical accuracy we
 cannot in fact distinguish quasiperiodic regimes from periodic 
ones with large period).

\begin{figure}
\includegraphics[angle=270,width=\columnwidth]{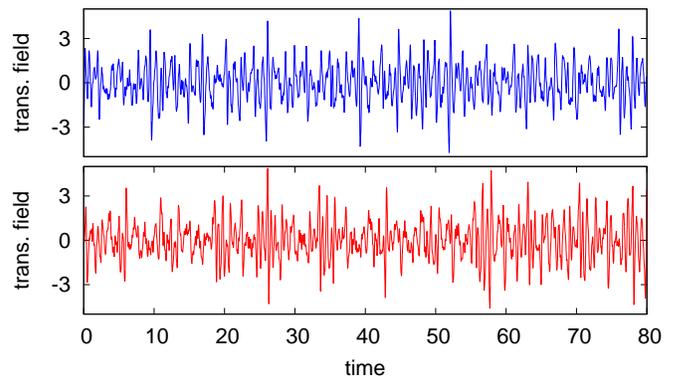}
\caption{Transmitted waves from left to right (bottom panel) and from right to left (top panel)
are chaotic. Time axis is
in units of the period of the incident wave.
}
\label{fig:chfield}
\end{figure}

For large amplitudes of incoming wave chaotic scattering in model
(\ref{eq:dde1}-\ref{eq:dde4}) is observed. We illustrate this in
Fig.~\ref{fig:chfield}, where we show transmitted fields for $A=20$ and
$\omega=1.45$. 

\begin{figure}
\includegraphics[width=\columnwidth]{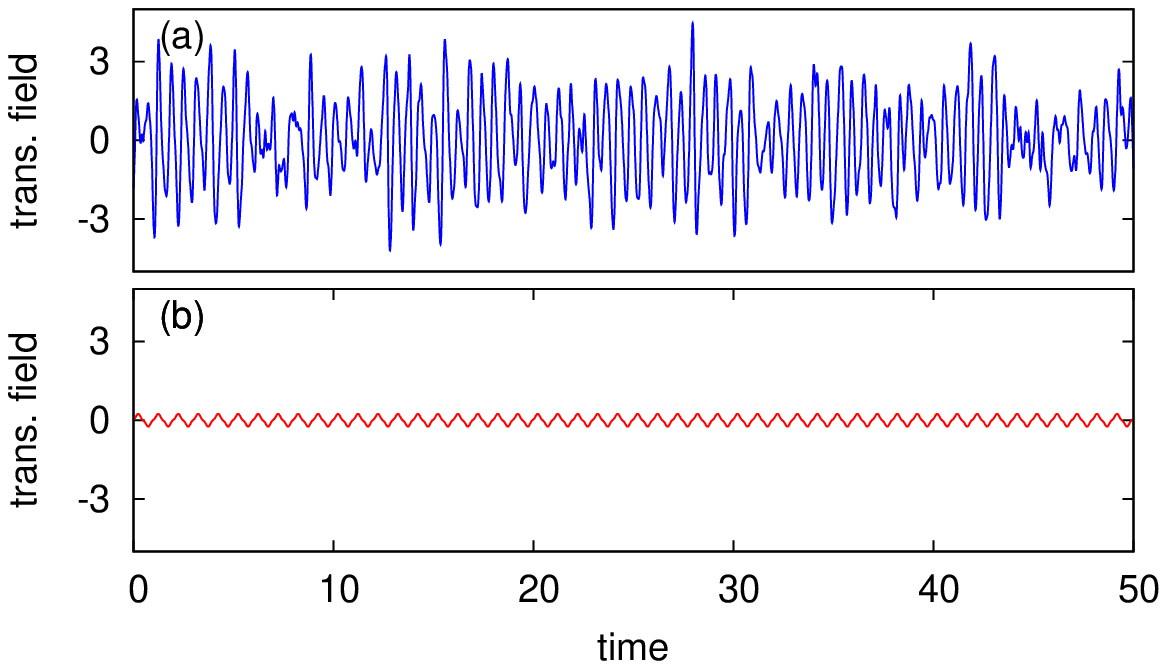}
\includegraphics[width=\columnwidth]{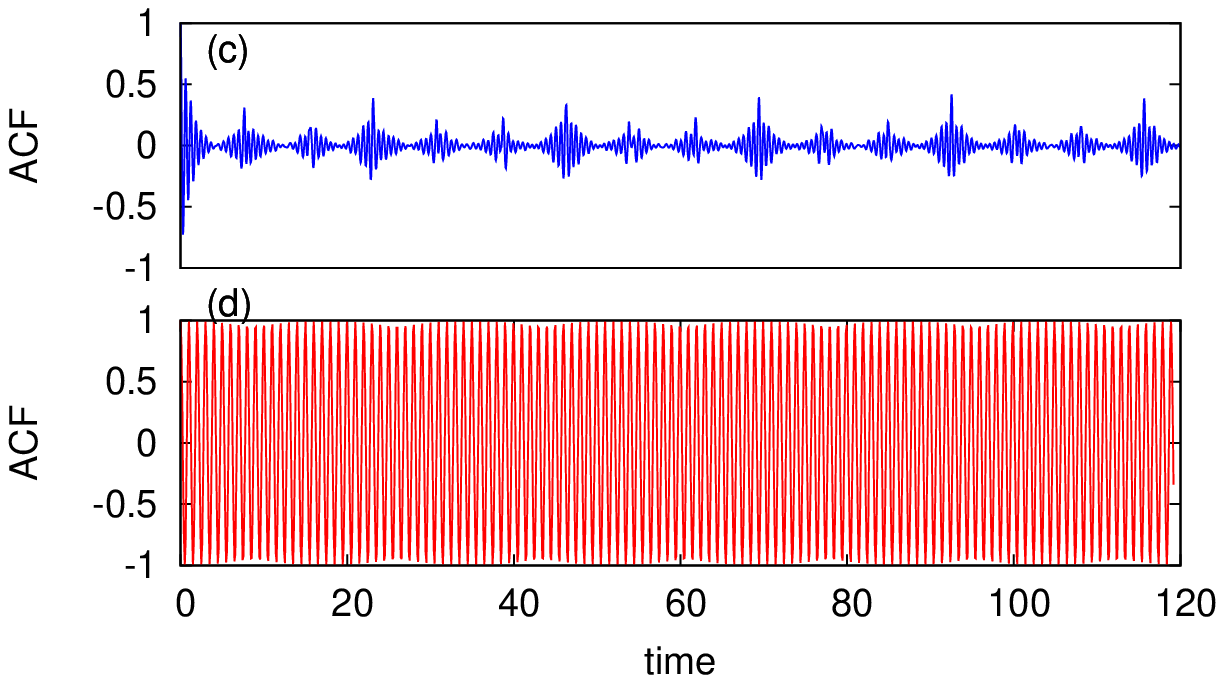}
\caption{Transmitted waves from left to right ((a), blue) and from right to 
left ((b), red), and their autocorrelation functions (panels (c,d)). 
Parameters: $A=16$, $\omega_1=
0.6$, $\omega_2=1.8$, $\omega=1.5$. Time axis is
in units of the period of the incident wave.}
\label{fig:chper}
\end{figure}

Probably, the mostly nontrivial situation is when the
transmission in one direction is chaotic, while in other direction periodic. 
We explored several sets of parameters and found
such a situation for the ``resonant'' frequencies of lumped oscillators
$\omega_2=1.8$, $\omega_1=0.6$. This ``chaotic diode'' regime is illustrated in 
Fig.~\ref{fig:chper}.

\section{Conclusions}

In this paper we described non-reciprocity effects in wave scattering on lumped nonlinear oscillators. We have analyzed equations describing a simple model of a linear string with two attached
oscillators, on two levels. Close to resonance we used amplitude equations, which allowed us 
a simplified analysis of
transmitted and reflected waves. Here we demonstrated non reciprocity 
and multistability of scattering. Already at his level of approximation,
we have shown that this setup can act as a nonreciprocal modulator via
a Hopf bifurcation of the steady solutions. 

In the second part, we performed a numerical analysis of full equations, and found more complex regimes of 
scattering: quasiperiodic and chaotic. A quite
interesting finding is that of chaotic non-reciprocity: while a periodic wave 
sent from one side remains periodic, the 
same wave sent on the system from the other side becomes 
chaotic. We think that such a regime might find application 
in chaotic communication. Unfortunately, we cannot link the two approaches.
In the first part the equations are derived in the asymptotic limit of large frequency, which
is hardly accessible in numerical studies of the full equations performed in part two. 
This is mainly due to neutral type of the appearing differential-delay equations. Another
difference is that in simulating the full equations we are not limited by a weak nonlinearity,
and in fact we considered rather large amplitudes to see chaotic regimes.

In most presented cases we reported scattering states obtained by direct numerical simulations. These yield only stable solutions. In several cases we revealed bistability: scanning solutions by slow change of frequency of the incident
wave different regimes have been obtained in some frequency ranges depending on whether it was decreased or increased. One cannot exclude higher degrees of multistability, i.e. co-existence of many stable branches, but such an analysis would
require much stronger computational efforts.

In the present work, we focused on an idealized system, where the waves are non-dispersive and there is no 
dissipation, neither in the wave propagation, nor in the lumped oscillators. This allowed us a coinsize formulation 
in terms of delayed differential equations, although of neutral type. For more realistic applications, e.g. in 
optical systems, one needs to incorporate effects of dispersion and diffusion/dissipation. 
We expect, however, that non-trivial 
regimes of complex non-reciprocity could be found 
in such systems as well.

\acknowledgments

We thank A. Politi for fruitful discussions. We acknowledge the Galileo Galilei Institute 
for Theoretical Physics (Florence, Italy) for the hospitality and the INFN for partial 
support during the completion of this work. The work of AP was partly supported by 
the grant (agreement 02.В.49.21.0003 of August 27, 2013  between the 
Russian Ministry of Education and Science and Lobachevsky State University of Nizhni Novgorod).

\appendix*
\section{Derivation of the equations}

In this Appendix we derive the equation of motion of the system.
We refer to Fig.~\ref{fig:model} and the main text for the definition of the various
quantities. For the string field, we have four boundary conditions
\begin{gather*}
v(t)=F(t)+\alpha(t-L/2c) = \nonumber \\
\phi(t+L/2c)+\psi(t-L/2c)\;,\\
u(t)=\beta(t-L/2c)=\phi(t-L/2c)+\psi(t+L/2c)\;, \nonumber
\end{gather*}
and the expressions for derivatives
\begin{gather*}
\frac{\partial y^0}{\partial x}=-\frac{1}{c}\dot \phi(t-x/c)+\frac{1}{c}\dot\psi(t+x/c)\;, \nonumber\\
\frac{\partial y^-}{\partial x}=-\frac{1}{c}\dot F(t-(x+L/2)/c)+\frac{1}{c}\dot\alpha(t+x/c)\;,\\
\frac{\partial y^+}{\partial x}=-\frac{1}{c}\dot \beta(t-x/c)\;. \nonumber
\end{gather*}
Substituting this in the equations for $v,u$ we get
\begin{gather*}
m_1\ddot v+V(v)= \\
= \frac{S}{c}(-\dot\phi(t+L/2c)+\dot\psi(t-L/2c)+\dot F(t)-\dot\alpha(t-L/c))\;, \nonumber \\
m_2\ddot u+U(u)=\\
= \frac{S}{c}(-\dot\beta(t-L/2c)-\dot\psi(t+L/2c)+\dot\phi(t-L/2c))\;. \nonumber
\end{gather*}
From the boundary conditions we can express $\alpha$ and $\beta$:
\[
\beta(t-L/2c)=u(t)\;,\qquad \alpha(t-L/2c)=v(t)-F(t)\;.
\]
Substitution of this gives
\begin{gather*}
m_1\ddot v+V(v)=\\
\frac{S}{c}(-\dot\phi(t+L/2c)+\dot\psi(t-L/2c)+\dot F(t)-\dot v(t)+\dot F(t))\;,\\
m_2\ddot u+U(u)=\frac{S}{c}(-\dot u(t)-\dot\psi(t+L/2c)+\dot\phi(t-L/2c))\;.
\end{gather*}
Furthermore, substituting
\begin{gather*}
\dot\psi(t+L/2c)=-\dot\phi(t-L/2c)+\dot u(t)\;,\\
\dot\phi(t+L/2c)=-\dot\psi(t-L/2c)+\dot v(t)\;,
\end{gather*}
yields the final system
\begin{gather*}
m_1\ddot v+V(v)=\frac{S}{c}(2\dot\psi(t-L/2c)+2\dot F(t)-2\dot v(t))\;,\\
m_2\ddot u+U(u)=\frac{S}{c}(-2\dot u(t)+2\dot\phi(t-L/2c))\;,\\
\dot\psi(t)=\dot u(t-L/2c)-\dot\phi(t-2L/2c)\;,\\
\dot\phi(t)=\dot v(t-L/2c)-\dot\psi(t-2L/2c)\;.
\end{gather*}

We introduce the time scale according to some frequency $\Omega$, so that the new dimensionless time will be $\tau=\Omega t$.
Furthermore, we restrict to the case $m_1=m_2=m$. Then
\begin{gather*}
\frac{d^2 v}{d\tau^2}+\frac{V(v)}{m\Omega^2}=\frac{S}{mc\Omega}(2\dot\psi(\tau-\Omega L/2c)+2\dot F(\Omega^{-1}\tau)-2\dot v)\;,\\
\frac{d^2 u}{d\tau^2}+\frac{U(u)}{m\Omega^2}=\frac{S}{mc\Omega}(-2\dot u+2\dot\phi(\tau-\Omega L/2c))\;,\\
\dot\psi(\tau)=\dot u(\tau-\Omega L/2c)-\dot\phi(\tau-2\Omega L/2c)\;,\\
\dot\phi(\tau)=\dot v(\tau-\Omega L/2c)-\dot\psi(\tau-2\Omega L/2c)\;.
\end{gather*}
which upon suitable parameter redefinition reduces to system (\ref{eq:dde1})-(\ref{eq:dde4}).

\bibliography{diodo,additionalrefs}
\bibliographystyle{apsrev}
\end{document}